\documentclass[4pt,aps,pre,reprint,superscriptaddress
]{revtex4-2} %
\usepackage[dvipsnames]{xcolor}
\usepackage[utf8]{inputenc}
\usepackage[T1]{fontenc}
\usepackage{amsmath,amssymb,amsthm,bm}

\usepackage{hyperref}

\usepackage{dcolumn}
\usepackage{mathptmx}
\usepackage{amsmath,amssymb,mathtools}
\usepackage{physics}
\usepackage{enumitem}
\usepackage[cal=dutchcal,scr=rsfs,frak=esstix,bb=ams]{mathalfa}

\begin{document}
\title{Criticality Beyond Nonanalyticity: Intrinsic Microcanonical Signatures of Phase Transitions}
\author{Loris Di Cairano}
\affiliation{Department of Physics and Materials Science, University of Luxembourg, L-1511 Luxembourg City, Luxembourg}
\email{l.di.cairano.92@gmail.com, loris.dicairano@uni.lu}


\date{\today}

\begin{abstract}
Phase transitions are conventionally defined by nonanalyticities of thermodynamic potentials in the thermodynamic limit. In this Letter, we show that the singularity is not the definition of criticality but its asymptotic outcome: criticality is already written in the microcanonical entropy derivatives at any finite size as intrinsic morphological structures---inflection points and extrema. The singularity is then the endpoint of a sharpening process that evolves with increasing system size.
Combining microcanonical inflection-point analysis (MIPA) with the Berlin-Kac spherical model---for which the microcanonical density of states is known in closed form at every finite $N$---we systematically identify these structures in the energy profiles of entropy derivatives that encode the transition. An inflection point in the inverse temperature $\beta_N(\epsilon)=\partial_\epsilon S_N$ and a pronounced peak in its derivative $\gamma_N(\epsilon)=\partial^2_\epsilon S_N$ define a well-controlled pseudocritical trajectory whose controlled sharpening and drift culminate in the macroscopic cusp at the critical energy $\epsilon_c$ in the thermodynamic limit. This establishes an intrinsic, order-parameter-free notion of criticality that precedes its singular asymptotic representation.
\end{abstract}

\maketitle

Phase transitions mark the emergence of collective order in many-body systems. The standard theoretical framework identifies them with nonanalyticities of thermodynamic potentials in the limit $N\to\infty$: a free energy that loses analyticity signals a transition; where it remains smooth, no transition is declared. Detecting the transition then requires an apparatus---order parameters, broken symmetries, correlation functions, finite-size scaling ans\"atze~\cite{brezin1988finite,fisher1972scaling,binder1981finite}---designed to reconstruct a singularity that, by construction, is never directly accessible.

Yet a singularity does not appear out of nowhere at $N=\infty$. Something in the thermodynamics at finite $N$ must already encode the information that will become the nonanalyticity. How this encoding works is answered differently by different frameworks. The renormalization group traces criticality to fixed points of a coarse-graining flow, requiring the identification of relevant operators and universality class. The Lee-Yang theory locates it in the migration of partition-function zeros toward the real axis~\cite{yang1952statistical,lee1952statistical}, a criterion tied to the canonical partition function. Finite-size scaling organizes the approach to the singularity through system-dependent ans\"atze built on the chosen order parameter~\cite{brezin1988finite,binder1981finite}. Each of these is powerful, but each reconstructs the transition indirectly---from the outside, through a formalism-specific lens.

Here we change perspective. Rather than reconstructing the singularity from auxiliary constructions, we identify it at its origin: directly in the microcanonical thermodynamic observables. The microcanonical ensemble provides the natural framework~\cite{gross2005microcanonical,gross2001microcanonical,gross2000phase,gross2001ensemble} since it is the only thermodynamic description that remains well-defined even in regimes where canonical descriptions become ambiguous or fail---a situation made rigorous by the theory of ensemble inequivalence~\cite{touchette2005nonequivalent,touchette2004introduction,touchette2003equivalence,ellis2002nonequivalent,ellis2004thermodynamic}. This is particularly relevant for long-range interacting 
systems~\cite{campa2014physics,barre2001inequivalence,leyvraz2002ensemble}, 
including regimes with negative absolute 
temperature~\cite{miceli2019statistical}, in quantum field theory~\cite{strominger1983microcanonical}, in gravitational systems~\cite{brown1993microcanonical}, and in the  thermodynamics of hadronic matter~\cite{Hagedorn1965}. 

Within this framework, we demonstrate that the energy derivatives $\beta_N(\varepsilon)=\partial_\varepsilon S_N$ and $\gamma_N(\varepsilon)=\partial_\varepsilon^2 S_N$ carry a morphological structure that is the direct generator of the macroscopic singularity. This structure consists of elementary objects---inflection points in $\beta_N$, peaked extrema in $\gamma_N$---that are few, classifiable, and the same at every $N$. Moreover, they do not depend on the model, on the choice of an order parameter, or on any scaling hypothesis. One identifies them at a given $N$ and already holds the object responsible for the thermodynamic nonanalyticity. Tracking their evolution at increasing $N$ then amounts to watching the singularity being built, step by step.

The microcanonical inflection-point analysis (MIPA)~\cite{qi2018classification,bachmann2014novel} provides the formal framework that classifies these structures and distinguishes physical signals from artifacts; it has been validated across spin models~\cite{sitarachu2020exact,sitarachu2022evidence,liu2025geometric,rocha2025microcanonical}, polymers~\cite{bachmann2014novel,schnabel2011microcanonical,koci2017subphase}, lattice field theories~\cite{bel2021geometrical,di2022geometrictheory}, glass-transitions~\cite{vesperini2025glass}, and long-range systems~\cite{di2025geometric,di2025phase}. What has been missing is the definitive analytical proof: a model where one can follow the inflection point from any finite $N$ all the way to the singularity, with full control over every step, and verify it independently against dynamical simulations.

\begin{figure*}
    \centering
    \includegraphics[width=0.8\linewidth]{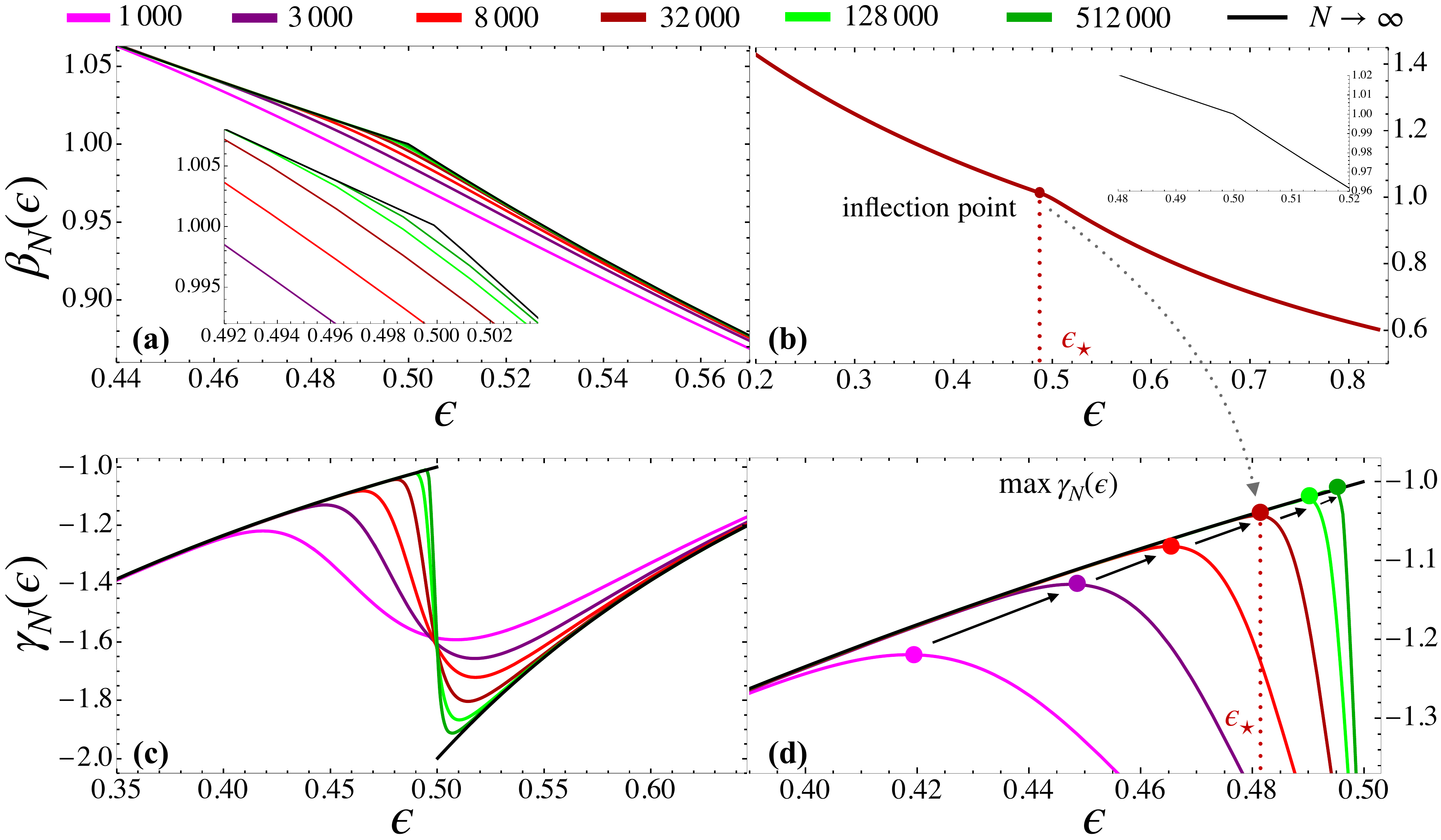}
    \caption{\textbf{Finite-size first- and second-order derivatives of entropy in the mean field Berlin-Kac model.}
Microcanonical inverse temperature $\beta_N(\varepsilon)=\partial_\varepsilon s_N(\varepsilon)$ (top) and its derivative
$\gamma_N(\varepsilon)=\partial_\varepsilon\beta_N(\varepsilon)=\partial_\varepsilon^2 s_N(\varepsilon)$ (bottom), shown for increasing system sizes $N$ (colored curves) together with the thermodynamic-limit prediction $N\to\infty$ (black curve).
\textbf{(a.1)} Family $\{\beta_N(\varepsilon)\}_N$ converging to the infinite-size curve, which exhibits a cusp at the critical energy $\varepsilon_c=J/2=1/2$ (inset highlights the sharpening near $\varepsilon_c$).
\textbf{(a.2)} A representative finite size (here $N=32000$) revealing an intrinsic inflection point in $\beta_N$ at $\varepsilon_\star$ (vertical guide).
\textbf{(b.1)} Corresponding family $\{\gamma_N(\varepsilon)\}_N$ showing the characteristic finite-$N$ ``Z-like'' structure in the critical region.
\textbf{(b.2)} Zoom on the negative peak of $\gamma_N$; its location $\varepsilon_\star(N)$ (colored disks) drifts toward $\varepsilon_c$ while the peak sharpens with $N$, demonstrating that the macroscopic cusp/discontinuity emerges as the singular endpoint of a fully trackable finite-$N$ microcanonical structure.}
    \label{fig:mipa}
\end{figure*}

In this Letter, we provide this proof. In the mean-field Berlin-Kac spherical model, the microcanonical density of states is known in closed form at every $N$~\cite{casetti2006nonanalyticities,kastner2006mean}, allowing us to compute $\beta_N(\varepsilon)$ and $\gamma_N(\varepsilon)$ exactly---and to compare them one-to-one with microcanonical molecular dynamics at the same $N$. What we find is a precise mechanism: $\beta_N(\varepsilon)$ develops an intrinsic inflection point, whose sharpening with $N$ produces the macroscopic cusp at the critical energy $\varepsilon_c$, while $\gamma_N(\varepsilon)$ exhibits a negative peak at $\varepsilon_\star(N)$ that drifts toward $\varepsilon_c$, and whose divergence generates the thermodynamic discontinuity. Because the density of states near $\varepsilon_c$ reduces to a Landau-type saddle~\cite{brezin1988finite,binder1985finite}, this drift is not phenomenological: it scales as $N^{-1/2}$ with analytically fixed corrections, and the singularity is recovered as its controlled outcome. MIPA predicts this morphology---inflection plus negative-valued maximum---as the admissible signature of a continuous transition, and the exact solution confirms it. The conventional framework is not contradicted; it is contained as the asymptotic projection of a more elementary, model-independent structure.


The thermodynamic potential of the microcanonical ensemble is the entropy function,
$S_N(E)=\ln\Omega_N(E)$, namely, the logarithm of the density of states (DoS)~\cite{pathria2017statistical}
\begin{equation}
    \Omega_N(E)=\int \delta\big(H(\bm{p},\bm{q})-E\big)\,d^Nq\,d^Np\,,
\end{equation}
where $H$ is the Hamiltonian of the system and $E$ is the fixed energy value. Entropy is an
intrinsic object that generates all thermodynamic observables through its energy derivatives, such as, for instance:
\begin{equation}
\label{def:thermo-observables}
\beta_N(E):=\frac{\partial S_N}{\partial E},
\qquad
\gamma_N(E) := \frac{\partial \beta_N}{\partial E}=\frac{\partial^2 S_N}{\partial E^2}.
\end{equation}
Here, $\beta_N\equiv 1/T$ corresponds to the inverse temperature. Our aim is precisely
to show that the transition is \emph{not silent} in these thermodynamic observables: at any finite
$N$, it leaves an intrinsic, morphological structure whose sharpening and drift culminate in the thermodynamic singularity. However, the main difficulty lies in identifying coherent rules that allow us to detect them and distinguish physical from nonphysical signals.

The mean-field (fully connected) Berlin-Kac spherical model, defined by the Hamiltonian with constraint~\cite{kastner2006mean,casetti2006nonanalyticities}
\begin{equation}
H(\bm{q},\bm{p})=\frac12\sum_{i=1}^N p_i^2-\frac{J}{2N}\bigg(\sum_{i=1}^N q_i\bigg)^2\,,
\qquad\sum_{i=1}^N q_i^2 = N,
\end{equation}
provides precisely such a testbed since the DoS that can be computed exactly in closed form for every finite $N$~\cite{casetti2006nonanalyticities}, namely:
\begin{widetext}
\begin{equation}
\begin{split}
    \Omega_N(\epsilon) &\propto \Gamma\left(\!\frac{N-1}{2}\!\right)\Gamma\left(\frac{N}{2}\right)
\,(1+2\epsilon)^{N-2}\;\;{}_2F_{1}\!\left(\dfrac{1}{2},\dfrac{N-1}{2};\,N-1;\,1+2\epsilon\right),\qquad -1/2<\epsilon\le 0,\\
    \Omega_N(\epsilon) &\propto \sqrt{\pi}(N-1)(2\epsilon)^{\frac{N-3}{2}}\;\;
{}_2F_{1}\!\left(\dfrac{1}{2},\dfrac{3-N}{2};\,\dfrac{N}{2};\,-\dfrac{1}{2\epsilon}\right)\,,\hspace{3.7cm} 0<\epsilon\,,
\end{split}
\end{equation}
\end{widetext}
where $J=1$ and $\epsilon:=E/N$ is the specific energy. Then, ${}_2F_{1}$ and $\Gamma$ denote the hypergeometric and Gamma functions, respectively. Note that here, the DoS must include a Dirac delta function $\delta(\sum_i q_i^2-N)$ due to the further constraint besides the fixed-energy constraint. We used this finite-$N$ solution to compute exactly the microcanonical observables defined in Eq.~\eqref{def:thermo-observables}. Any technical detail for this calculation is reported in Sec.~S1 of the Supplemental Material (SM). The panels~\ref{fig:mipa}\textbf{(a)} and \ref{fig:mipa}\textbf{(b)} show the inverse temperature function $\partial_\epsilon s(\epsilon)\equiv\beta_N(\epsilon)$ for increasing system sizes where we used the specific entropy $s(\epsilon)=S_N(\epsilon)/N$. In panel~\ref{fig:mipa}\textbf{(a)}, we show the family $\{\beta_N(\epsilon)\}_N$ (colored curves) converging to its infinite-size curve (the black dashed curve); the latter displays a cusp (see the inset) at the critical energy $\epsilon_c(\infty)=J/2=1/2$ as known in the literature~\cite{casetti2006nonanalyticities}. During this convergence process, the $\beta_N$-functions change their energy profile in a way that is quite subtle. Therefore, in order to better visualize such a change, we report in panel~\ref{fig:mipa}\textbf{(b)} only one member of this family ($N=32\,000$). In doing so, we notice the presence of an inflection point at $\epsilon_\star=0.482$. From a simple argument in real analysis, we know that if a function $x\mapsto f(x)$ changes from upward to downward concavity, then an inflection point is present, and this gives rise to a maximum in its derivative $f'(x)$. This is confirmed by the presence of a (negative-valued) maximum in $\gamma_N\equiv\partial_E\beta_N$ at $\epsilon_\star=0.482$, as we see in panel \textbf{(d)}. 

Remarkably, by inspecting Fig.~\ref{fig:mipa}\textbf{(c)}, we observe that, at every system size, the respective $\gamma_N$-function displays a maximum; therefore, we can conclude that this structural feature is not just relegated to a specific system size, but is an intrinsic property of the entire family $\{\beta_N,\gamma_N\}_N$. However, the size affects the energy location of the maximum. We then define the energy values associated with the maximum of $\gamma_N$:
\begin{equation*}
    \epsilon_\star(N):\;\;\;\Big.\gamma_N(\epsilon_\star(N)) \ \text{is a local max with}\;\gamma_N(\epsilon_\star)<0\,,
\end{equation*}
and the value of the maximum
\begin{equation*}
    M(N):= \gamma_N\big(\epsilon_\star(N)\big)\equiv\max_\epsilon\gamma_N(\epsilon)\,.
\end{equation*}
Fig.~\ref{fig:mipa}\textbf{(d)} shows that the smaller the size, the lower the energy and maximum values; namely, for $N<N^\prime$, we have $\epsilon_\star(N)<\epsilon_\star(N^\prime)$ and $M(N)<M(N^\prime)$. Interestingly, as the size increases, the energy profile of $\gamma_N$ sharpens, and the sequences $\{\epsilon_\star(N)\}_N$ and $\{M(N)\}_N$ simultaneously drift with $N$ thus reaching the asymptotic value $\lim_{N\to\infty}\epsilon_\star(N)=\epsilon_c=0.5$ and $\lim_{N\to\infty}M(N)=-1\equiv\gamma_\infty(\epsilon_c)$. This process is represented by the sequences of the colored disks in panel~\ref{fig:mipa}\textbf{(d)}.

In conclusion, the inflection points are the structures that transform into a cusp in the thermodynamic limit at the level of the inverse temperature $\beta_N$, while the maxima in $\gamma_N$ are the markers whose sharpening produces the thermodynamic divergence.

\begin{figure}
    \centering
    \includegraphics[width=1\linewidth]{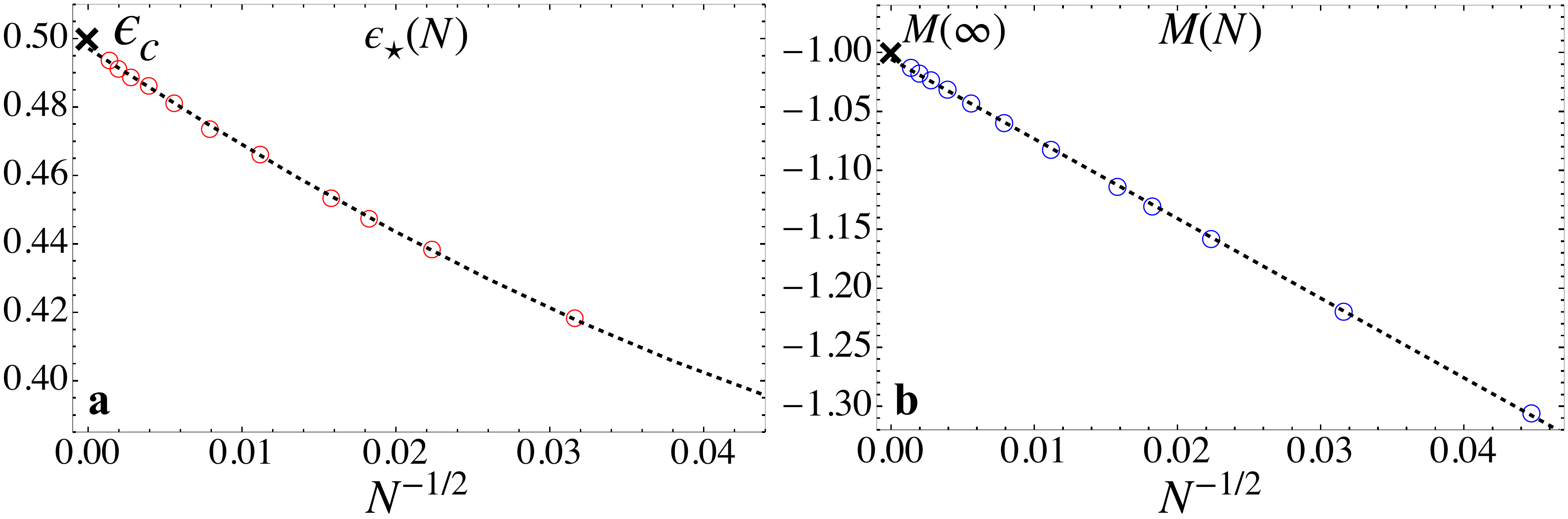}
    \caption{\textbf{Finite-size drift of the microcanonical precursor toward the critical energy.}
    \textbf{(a)} Pseudo-critical energy $\varepsilon_\star(N)$ defined as the location of the negative peak of $\gamma_N(\varepsilon)=\partial_\varepsilon^2 s_N(\varepsilon)$, plotted versus $N^{-1/2}$.
    The fit $\varepsilon_\star(N)=a+b\,N^{-1/2}+c\,N^{-1}$ yields $a\simeq0.4973$, consistent with $\varepsilon_\star(N)\to \varepsilon_c=1/2$ as $N\to\infty$.
    \textbf{(b)} Peak height $M(N)=\gamma_N(\varepsilon_\star(N))$ approaching the thermodynamic-limit value $M(\infty)=-1$.}
    \label{fig:scaling}
\end{figure}

Having established that the thermodynamic cusp and divergence arise as the singular limit of a
well-defined finite-$N$ structure, the next step is to quantify \emph{how} this structure approaches its asymptotic form with increasing size. In our setting, this is not an \emph{ad hoc} finite-size ansatz: for the Berlin-Kac model the approach to criticality is constrained by the analytic structure of the DoS itself. Close to the critical energy $\epsilon_c$, the large-$N$ exponent admits a Landau-like expansion governed by the symmetry-breaking pitchfork (see Sec.~S2 of the SM),
\begin{equation}
\Omega_N(\epsilon) \approx \int dm\ g(m)\,
\exp\!\left\{N\Big[f_0(\epsilon)+a(\epsilon)m^2-bm^4\Big]\right\},
\label{eq:landau_microcanonical}
\end{equation}
with $a(\epsilon)\simeq c(\epsilon-\epsilon_c)$ and $b>0$. Thus, the finite-size critical behavior is controlled only by the first stabilizing nonlinearity ($m^4$) and is universal within the mean-field (Landau) class~\cite{binder1981finite,brezin1988finite}. This structure implies a single scaling variable
\begin{equation}
    u \equiv (\epsilon-\epsilon_c)\,N^{1/2},
\end{equation}
obtained by the standard rescaling $m=N^{-1/4}y$ in Eq.~\eqref{eq:landau_microcanonical}. This is a scaling regime analogous to one identified in canonical finite-size analyses~\cite{binder1985finite,binder1981finite}, but derived directly from the microcanonical DoS.
Consequently, any \emph{local} pseudo-critical prescription based on a distinguished feature of the microcanonical derivatives---in particular the extremum of $\gamma_N$ used above---selects an $O(1)$ value $u=u_\star$, and therefore predicts the asymptotic shift
$\epsilon_\star(N)-\epsilon_c \sim u_\star\,N^{-1/2}$.

In practice, accessible sizes probe a pre-asymptotic regime where subleading corrections cannot be neglected. As shown in Sec.~S3 of the SM, the microcanonical scaling form obtained from
Eq.~\eqref{eq:landau_microcanonical} yields the controlled expansions
\begin{equation}
\begin{split}
    \epsilon_\star(N)&=\epsilon_c + b\,N^{-1/2} + c\,N^{-1/2-\omega}+\cdots,
    \\
    M(N)&=B\,N^{\beta}\big[1+O(N^{-\omega})\big],
\end{split}
\end{equation}
which we use below to extrapolate $\epsilon_c$ and quantify the sharpening of the finite-$N$
critical precursor.

\begin{figure}
    \centering
    \includegraphics[width=0.7\linewidth]{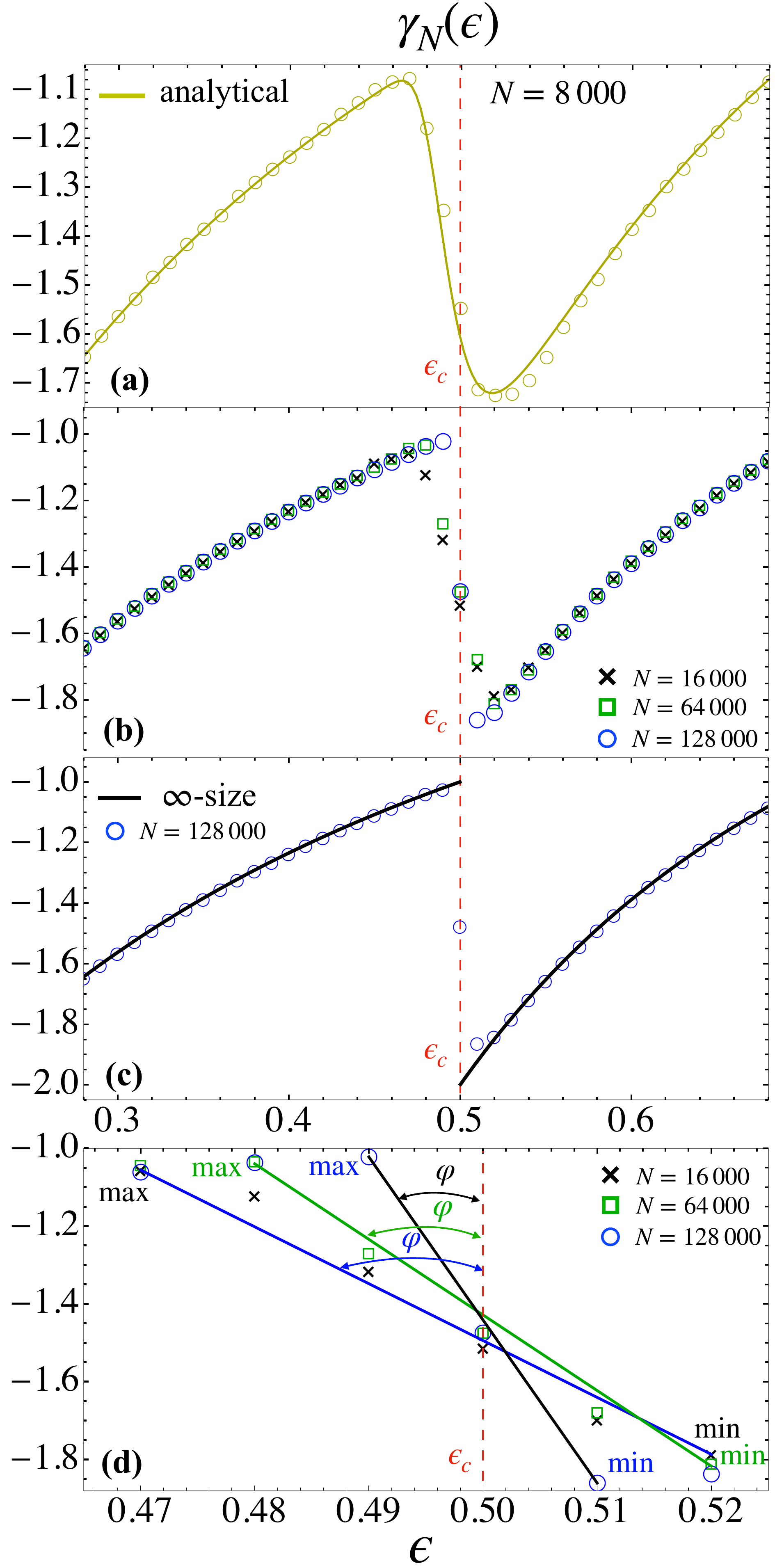}
    \caption{\textbf{Validation of the MIPA from numerical simulations.} Numerical $\gamma_N(\varepsilon)=\partial_\varepsilon^2 s_N(\varepsilon)$ from microcanonical molecular dynamics compared with exact analytical curves. \textbf{(a)} Simulated $\gamma_N(\varepsilon)$ for $N=16\,000$ (crosses), $N=64\,000$ (squares), and $N=128\,000$ (circles), showing the max--min structure around $\varepsilon_c$ (red dashed line). \textbf{(b)} Detail of the negative peak region; the rounded extremum is the MIPA second-order marker. \textbf{(c)} Overlay with the thermodynamic-limit curve (black), showing systematic sharpening toward the nonanalytic profile. \textbf{(d)} Geometric quantification: the secant connecting local max and min near $\varepsilon_c$ steepens with $N$ (increasing angle $\varphi$), measuring convergence to the singularity.}
    \label{fig:numerical-second-deriv}
\end{figure}

Figure~\ref{fig:scaling} turns the qualitative ``drift-and-sharpening'' scenario of Fig.~\ref{fig:mipa} into a controlled extrapolation. Since the Landau-type saddle in Eq.~\eqref{eq:landau_microcanonical} leaves a single scaling variable $u=(\epsilon-\epsilon_c)N^{1/2}$, the marker sequence $\epsilon_\star(N)$ must approach $\epsilon_c$ with a leading $N^{-1/2}$ shift. The first correction is analytic: the next stabilizing term in the exponent is $m^6$, and under $m=N^{-1/4}y$ it contributes as $N m^6\sim N^{-1/2}$, i.e., $\omega=1/2$ and a $1/N$ correction; hence, the next term scales as $N^{-1}$.
Accordingly, panel~\textbf{(a)} is accurately described by
$\epsilon_\star(N)=a+bN^{-1/2}+cN^{-1}$, yielding $a=0.497322$ and confirming convergence to $\epsilon_c=0.5$ from below ($b<0$). Panel~\textbf{(b)} shows that the corresponding peak value $M(N)$ approaches its asymptotic limit smoothly, completing the picture: the thermodynamic cusp and divergence are the singular endpoint of a finite, fully trackable microcanonical trajectory.

At this stage, we first focus on observing the microcanonical observables themselves. The appearance of inflection points in $\beta_N(\varepsilon)$ and of corresponding extrema in $\gamma_N(\varepsilon)=\partial_\varepsilon \beta_N$ is not an accidental feature of the Berlin-Kac model, nor a mere finite-size rounding: it is the generic local structure through which microcanonical thermodynamics can encode a collective reorganization at finite $N$. These are precisely the only admissible finite-$N$ markers of criticality: cooperative behavior is detected by least-sensitive inflection structures of the entropy or its derivatives. These features have been observed and deeply investigated from different perspectives for a long time~\cite{gross2005microcanonical,gross2002geometric,gross2001ensemble,pleimling2005microcanonical,behringer2006continuous} and are usually thought to exist only for finite-size systems \cite{chomaz2006challenges,chomaz1999energy,gulminelli1999critical}. However, the introduction of the microcanonical inflection-point analysis (MIPA)~\cite{qi2018classification,bachmann2014novel}, based on the principle of minimal sensitivity~\cite{stevenson1981optimized,stevenson1981resolution}, allows us to extend these approaches for classifying phase transitions also in the thermodynamic limit. Focusing on second-order transitions, these occur if and only if $\beta_N$ admits an inflection point in $\epsilon_\star$ such that $\gamma_N$ admits a negative-valued maximum; other pronounced extrema (e.g., the positive-valued minima) are not classified as transitions. A coherent generalization of the definition of phase transition with higher-than-second-order is also provided by MIPA~\cite{qi2018classification}. Now, we demonstrate the validity of MIPA in the thermodynamic limit in mean-field regimes (Berlin-Kac). We stress that we do not define finite-$N$ transitions via nonanalyticities. The finite-$N$ microcanonical entropy may display nonanalyticities~\cite{hilbert2014thermodynamic,dunkel2006phase} that do not survive the thermodynamic limit~\cite{casetti2006nonanalyticities,CasettiKastnerNerattini2009}.

Having identified which morphological structures of entropy are admissible markers of criticality, the real question now becomes practical: what can be done when an exact solution is \emph{not} available and one only has data from finite-size simulations? Should rounded extrema of entropy derivatives be dismissed as a mere crossover?

At finite $N$, microcanonical observables from simulations are unavoidably smooth, so the critical region can easily be misread as a featureless crossover. It is precisely here that one risks discarding genuine information by labeling such features as mere crossover. The point of the next comparison is to show that this is not justified. For the Berlin-Kac model we can place the numerical estimates and the exact finite-$N$ curves on the same footing and, at the same time, benchmark their evolution against the thermodynamic-limit reference. This allows us to demonstrate that the seemingly innocuous smooth morphology in $\gamma_N(\varepsilon)$ is a \emph{reliable} signature: it is reproduced quantitatively by simulations at the same $N$ and it sharpens systematically toward the nonanalytic limiting profile as $N$ increases.
This statement is then proved in Fig.~\ref{fig:numerical-second-deriv} in a way that remains meaningful even in a model without an exact solution: at finite $N$ the microcanonical signal is \emph{necessarily} rounded. It is exactly this rounded morphology that carries the information about the emerging transition.

Panel~\ref{fig:numerical-second-deriv}\textbf{(a)} shows the situation one would normally face in a model without an exact solution. For the very small size, (here, $N=8000$), numerical $\gamma_N(\varepsilon)$ exhibits a smooth rounded peak in the critical window (see sand-colored circles). This is the kind of feature that is conventionally classified as a crossover. The key observation is that here we can check this interpretation: those same points fall directly on the exact finite-$N$ curve (sand-colored solid line). Therefore, the rounded peak is not spurious; it is the correct microcanonical observable at that $N$. Discarding it would mean discarding genuine physics.

Panel~\ref{fig:numerical-second-deriv}\textbf{(b)} then shows what changes with system size. The same peak does not disappear; instead, it sharpens systematically as $N$ increases, while remaining confined to the same narrow energy region around $\varepsilon_c$ (red dashed line). In other words, the finite-size effect is not the creation of a new structure, but the progressive sharpening of an already-present one.

Panel~\ref{fig:numerical-second-deriv}\textbf{(c)} makes the connection to the macroscopic regime explicit: at the larger size $N=128\,000$, the numerical curve lies extremely close to the thermodynamic-limit prediction, displaying the same sharpening trend toward the nonanalytic profile. Finally, panel~\ref{fig:numerical-second-deriv}\textbf{(d)} converts ``sharper'' into a geometric statement: the secant connecting the local maximum and minimum becomes progressively steeper (increasing angle $\varphi$) with $N$, providing a quantitative measure of the approach to the singular limit.


In this Letter, we have shown that the standard 
identification of phase transitions with nonanalyticities 
in the thermodynamic limit is not a definition of 
criticality, but one particular representation of it. The 
singularity does not need to be reconstructed from auxiliary constructions---order parameters, symmetry-breaking patterns, scaling ans\"atze---because its origin is directly visible in the microcanonical thermodynamic observables: an inflection point in the inverse temperature 
$\beta_N(\varepsilon)$ and the associated negative peak in 
$\gamma_N(\varepsilon)$, present at every $N$ and classifiable through the microcanonical inflection-point analysis. The method is entirely agnostic: it requires no prior identification of an order parameter, no assumption about symmetry breaking, no scaling hypothesis, and no reference to a canonical ensemble. The entropy derivatives already carry the full signature of the transition at every $N$; the only task is to 
identify the correct morphology---which MIPA provides---rather than to reconstruct a singularity that is never directly accessible.

In the Berlin-Kac spherical model, where the full finite-$N$ microcanonical thermodynamics is available analytically, we have closed the logical loop: the morphological structure is computed exactly, reproduced by microcanonical molecular dynamics at the same $N$, and its controlled drift and sharpening---governed by the Landau saddle structure of the density of states---culminate in the macroscopic cusp at $\varepsilon_c$. The same morphology has been identified in the 1D XY model in both mean-field and long-range ($1/r^\alpha$) regimes~\cite{di2025geometric,di2025phase}, 
confirming that it is not model-specific.

The operational consequence is immediate: rounded extrema of entropy derivatives in finite-size simulations should not be dismissed as crossover. They are the transition---read at finite $N$ rather than inferred from an inaccessible limit. 
This makes criticality accessible in regimes where canonical diagnostics are ambiguous or fail: long-range interactions with ensemble inequivalence, gravitational systems, quantum field theories at fixed energy, and hadronic matter near the Hagedorn temperature. More broadly, the result establishes that the conventional thermodynamic framework---finite-size scaling, critical exponents, universality classes---is not contradicted but \emph{contained}: it emerges as the asymptotic projection of a more elementary, model-independent structure that is already fully operative at every finite $N$.

The calculations presented in this paper were carried out using the HPC facilities of the University of Luxembourg~\cite{VBCG_HPCS14} {\small (see \href{http://hpc.uni.lu}{hpc.uni.lu})} and those of the Luxembourg
national supercomputer MeluXina.

The code used for the numerical simulations is publicly available at \href{https://github.com/ldicairano/Berlin\_Kac\_MeanField}{github.com/Berlin\_Kac\_MeanField}.

\clearpage
\onecolumngrid
\begin{center}\textbf{\Large Supplemental Material\\
Criticality Beyond Nonanalyticity: Intrinsic Microcanonical Signatures of Phase Transitions}\end{center}

\setcounter{secnumdepth}{2} 

\renewcommand{\thesection}{S\arabic{section}}
\renewcommand{\thesubsection}{\arabic{subsection}}
\renewcommand{\thesubsubsection}{S\arabic{section}.\arabic{subsection}.\arabic{subsubsection}}

\renewcommand{\theequation}{S\arabic{equation}}
\renewcommand{\thefigure}{S\arabic{figure}}
\renewcommand{\thetable}{S\arabic{table}}

\setcounter{section}{0}
\setcounter{equation}{0}
\setcounter{figure}{0}
\setcounter{table}{0}

\section{Semi-analytic evaluation of finite-$N$ microcanonical derivatives}
\label{SM:hypergeom}

This section documents the semi-analytic pipeline used to compute finite-$N$ derivatives of the
microcanonical entropy density from an explicit expression of the density of states.
All numerical results in this part are produced by the Python script \texttt{omega\_hypergeom\_ode.py}, which evaluates $\Omega_N(\varepsilon)$ with arbitrary precision arithmetic and computes derivatives by finite differences. 

\subsection{Code availability}
\label{SM:code_hypergeom}

The script used for the semi-analytic evaluation described in this section is included in the
repository accompanying this work, and can be
run as a standalone program. The simulation code is publicly available 
\href{https://github.com/ldicairano/Berlin_Kac_MeanField/tree/main/Python}{in the GitHub repository}.

\subsection{Finite-$N$ density of states and entropy density}
\label{SM:dos_entropy}

For the Berlin--Kac mean-field spherical model one can write the finite-$N$ density of states
$\Omega_N(\varepsilon)$ (up to an overall multiplicative constant independent of $\varepsilon$)
in terms of a Gauss hypergeometric function. In the parametrization used here,
\begin{equation}
\Omega_N(\varepsilon)=
\frac{(2\varepsilon)^{\frac{N-1}{2}}}{1+2\varepsilon}\;
{}_2F_1\!\left(1,\,N-\frac{3}{2};\,\frac{N}{2};\,u(\varepsilon)\right),
\qquad
u(\varepsilon)=\frac{1}{1+2\varepsilon},
\qquad \varepsilon>0.
\label{SM:Omega_hypergeom}
\end{equation}
We define the finite-$N$ entropy density as
\begin{equation}
s_N(\varepsilon)\equiv \frac{1}{N}\ln \Omega_N(\varepsilon),
\label{SM:sN_def}
\end{equation}
and we are primarily interested in its derivatives
\begin{equation}
\beta_N(\varepsilon)\equiv \partial_\varepsilon s_N(\varepsilon),
\qquad
\gamma_N(\varepsilon)\equiv \partial_\varepsilon^2 s_N(\varepsilon),
\qquad \ldots
\end{equation}
The script computes $\beta_N$ and (optionally) $\gamma_N$ on a uniform grid in $\varepsilon$.

\subsection{Robust evaluation of $\ln \Omega_N(\varepsilon)$ at large $N$}
\label{SM:robust_logOmega}

Direct evaluation of ${}_2F_1(a,b;c;u)$ becomes numerically delicate for large $N$, because the
parameters
\begin{equation}
a=1,\qquad b=N-\frac{3}{2},\qquad c=\frac{N}{2},
\end{equation}
are large while the argument $u(\varepsilon)\in(0,1)$ may approach $1$ when $\varepsilon$ is small.
To maintain stability we evaluate \emph{the logarithm} of the density of states,
\begin{equation}
\ln \Omega_N(\varepsilon)=
\underbrace{\frac{N-1}{2}\ln(2\varepsilon)-\ln(1+2\varepsilon)}_{\text{prefactor}}
+\ln\!\left[{}_2F_1\!\left(1,\,N-\frac{3}{2};\,\frac{N}{2};\,u(\varepsilon)\right)\right],
\label{SM:logOmega_split}
\end{equation}
using arbitrary precision arithmetic (mpmath). The working precision is increased with $N$
according to a heuristic rule $d(N)$ (hundreds to $\sim 10^3$ decimal digits at the largest sizes)
to avoid loss of significance and spurious \texttt{NaN} values. 

\paragraph{Hypergeometric stabilization by analytic continuation.}
When the direct series-based evaluation of ${}_2F_1$ fails to converge within a large cutoff on
the number of terms, we use a standard analytic continuation identity that maps the argument from
$u\in(0,1)$ to
\begin{equation}
u'=\frac{u}{u-1}\in(-\infty,0),
\end{equation}
which is typically more stable for large parameters. In logarithmic form this yields
\begin{equation}
\ln\!\left[{}_2F_1(a,b;c;u)\right]
=
-a\,\ln(1-u)+\ln\!\left[{}_2F_1\!\left(a,\,c-b;\,c;\,\frac{u}{u-1}\right)\right],
\label{SM:2F1_transform}
\end{equation}
and the script evaluates the right-hand side at the same working precision $d(N)$. 
(When needed, the maximum number of series terms is further increased.)

\paragraph{Caching.}
Because finite-difference derivatives require repeated evaluations at nearby energies
$\varepsilon\pm h$, the computation caches $\ln\Omega_N(\varepsilon)$ at fixed $(N,d(N))$.
To make caching robust against floating-point roundoff, energies are rounded to a fixed number of
decimal digits before being used as cache keys.

\subsection{Finite-difference derivatives}
\label{SM:finite_diff}

Derivatives of $s_N(\varepsilon)$ are computed by centered finite differences on a uniform grid
$\{\varepsilon_i\}_{i=1}^{n_{\mathrm{pts}}}$, with $\varepsilon_i\in[\varepsilon_{\min},\varepsilon_{\max}]$.
Given a stepsize $h>0$, the first derivative is approximated by
\begin{equation}
\beta_N(\varepsilon)\equiv s_N'(\varepsilon)
\;\approx\;
\frac{s_N(\varepsilon+h)-s_N(\varepsilon-h)}{2h}.
\label{SM:fd_s1}
\end{equation}
Optionally, the second derivative is computed as
\begin{equation}
\gamma_N(\varepsilon)\equiv s_N''(\varepsilon)
\;\approx\;
\frac{s_N(\varepsilon+h)-2s_N(\varepsilon)+s_N(\varepsilon-h)}{h^2}.
\label{SM:fd_s2}
\end{equation}
In practice, distinct steps $h_1$ and $h_2$ may be used for \eqref{SM:fd_s1} and \eqref{SM:fd_s2}
to balance truncation error and numerical noise at large $N$. 

\subsection{Thermodynamic-limit reference curves}
\label{SM:asymptotic_ref}

For comparison, we also compute the thermodynamic-limit inverse temperature $\beta(\varepsilon)$
in the same energy range using its known piecewise expression (here for coupling $J=1$):
\begin{equation}
\beta(\varepsilon)=
\begin{cases}
\dfrac{1}{\varepsilon+\frac{J}{2}}, & -\dfrac{J}{2}\le \varepsilon \le \dfrac{J}{2},\\[2mm]
\dfrac{1}{2\varepsilon}, & \varepsilon\ge \dfrac{J}{2},
\end{cases}
\qquad (J=1).
\label{SM:beta_asym}
\end{equation}
The script exports this asymptotic curve on the same $\varepsilon$ grid to facilitate direct
overlay with finite-$N$ results. 
(An analogous expression for $\beta'(\varepsilon)=s''(\varepsilon)$ is implemented but can be
enabled/disabled depending on the plotting needs.)

\subsection{Exported numerical data and figures}
\label{SM:outputs}

All outputs are written to a dedicated \texttt{out/} directory in purely numeric two-column format
(without headers): the first column is $\varepsilon$ and the second is the corresponding quantity. 
Specifically, the script produces:
\begin{itemize}
  \item \texttt{out/s1\_asymptotic.dat}: $(\varepsilon,\beta(\varepsilon))$ for the thermodynamic-limit curve.
  \item \texttt{out/s1\_raw\_N\{N\}.dat}: $(\varepsilon,\beta_N(\varepsilon))$ computed by \eqref{SM:fd_s1}.
  \item \texttt{out/s1\_mapped\_N\{N\}.dat}: a linearly rescaled version of $\beta_N(\varepsilon)$
        mapped into the value range of the asymptotic curve on the chosen interval.
        This is used only as a visualization aid when the raw finite-$N$ curve is affected by
        an additive/multiplicative normalization mismatch in $\Omega_N$ that is irrelevant for
        identifying robust geometric features (inflection points and extrema of derivatives).
  \item \texttt{out/plot\_s1\_raw.pdf} and \texttt{out/plot\_s1\_raw.png}: overlay of $\beta_N(\varepsilon)$ (raw)
        for the simulated sizes and $\beta(\varepsilon)$.
  \item \texttt{out/plot\_s1\_mapped.pdf} and \texttt{out/plot\_s1\_mapped.png}: same overlay for the mapped curves.
\end{itemize}
The script contains the analogous machinery for second derivatives ($s''_N$) and the corresponding
plots and data files; these can be activated by enabling the relevant sections in the code.

\section{From the exact finite-$N$ density of states to a Landau-type integral over $m$}
\label{sec:supp_landau_from_exact}

We consider the kinetic mean-field spherical model with Hamiltonian
\begin{equation}
H_N(q,p)=T_N(p)+V_N(q),\qquad
T_N=\frac12\sum_{i=1}^N p_i^2,\qquad
V_N(q)=-\frac{1}{2N}\Big(\sum_{i=1}^Nq_i\Big)^2,
\end{equation}
with the spherical constraint $\sum_iq_i^2=N$ and the tangency constraint
$\sum_iq_ip_i=0$. The microcanonical density of states at energy density
$\epsilon=E/N$ can be written as the phase-space integral (up to normalization)~\cite{casetti2006nonanalyticities}
\begin{equation}
\Omega_N(\epsilon)=a_N\int_{\mathbb R^N}dq\int_{\mathbb R^N}dp\,
\delta\!\left(\sum_{i=1}^Nq_i^2-N\right)\,
\delta\!\left(\sum_{i=1}^Nq_ip_i\right)\,
\delta\!\left(T_N+V_N-N\epsilon\right).
\label{eq:Omega_phase_space}
\end{equation}

\subsection{Identification of the magnetization variable}
Introduce the (continuous) magnetization per degree of freedom
\begin{equation}
m(q)\equiv \frac{1}{N}\sum_{i=1}^Nq_i.
\label{eq:m_def}
\end{equation}
A convenient way to make $m$ explicit is to rotate coordinates so that the first axis is the
normalized uniform vector $e_1=(1,1,\dots,1)/\sqrt{N}$. In these coordinates the projection of
$q$ along $e_1$ is $q\cdot e_1=\sum_iq_i/\sqrt{N}=\sqrt{N}\,m$, hence on the
sphere $|q|=\sqrt{N}$ one may parametrize
\begin{equation}
q\cdot e_1 = |q|\cos\vartheta = \sqrt{N}\cos\vartheta
\quad\Longrightarrow\quad
m=\cos\vartheta,\qquad m^2=\cos^2\vartheta.
\label{eq:m_cos_theta}
\end{equation}
With this identification the potential energy becomes $V_N=-(N/2)m^2$.

\subsection{Exact integral representation as an integral over $m$}
Carrying out the spherical-coordinate manipulations described in Ref.~\cite{kastner2006mean},
one arrives at the exact representation
\begin{equation}
\Omega_N(\epsilon)\ \propto\ \int_0^1 dy\; y^{-1/2}(1-y)^{(N-3)/2}\,(2\epsilon+y)^{(N-3)/2}\,
\Theta(2\epsilon+y),
\label{eq:Omega_y_exact}
\end{equation}
where $\Theta$ is the Heaviside function. 
Using the identification $y=m^2$ from Eq.~\eqref{eq:m_cos_theta} and the change of variables
$y=m^2$ ($dy=2m\,dm$), we obtain $y^{-1/2}\,dy=2\,dm$ and therefore
\begin{align}
\Omega_N(\epsilon)
&\propto \int_0^1 dm\; (1-m^2)^{(N-3)/2}\,(2\epsilon+m^2)^{(N-3)/2}\,\Theta(2\epsilon+m^2)
\nonumber\\
&\propto \int_{-1}^{1} dm\; g(m)\,
\exp\!\left\{\frac{N-3}{2}\Big[\ln(1-m^2)+\ln(2\epsilon+m^2)\Big]\right\}\,
\Theta(2\epsilon+m^2),
\label{eq:Omega_m_exact}
\end{align}
where we used evenness in $m$ to extend the integration domain and absorbed
$O(1)$ prefactors into a smooth function $g(m)$ (constant to leading order here).

Equation~\eqref{eq:Omega_m_exact} is already of the Laplace form
$\Omega_N(\epsilon)\propto\int dm\; g(m)\exp\{N f(\epsilon,m)\}$, with
\begin{equation}
f(\epsilon,m)=\frac12\Big[\ln(1-m^2)+\ln(2\epsilon+m^2)\Big]+O(1/N).
\label{eq:f_def}
\end{equation}

\subsection{Landau expansion and mean-field critical point}
For $|m|\ll 1$ one expands $f(\epsilon,m)$ at fixed $\epsilon>0$:
\begin{align}
\ln(1-m^2) &= -m^2-\frac12 m^4+O(m^6),\\
\ln(2\epsilon+m^2) &= \ln(2\epsilon)+\frac{m^2}{2\epsilon}-\frac{m^4}{8\epsilon^2}+O(m^6).
\end{align}
Inserting into Eq.~\eqref{eq:f_def} yields
\begin{equation}
f(\epsilon,m)
= f_0(\epsilon)+a(\epsilon)\,m^2-b(\epsilon)\,m^4+O(m^6),
\label{eq:landau_f}
\end{equation}
with
\begin{equation}
f_0(\epsilon)=\frac12\ln(2\epsilon),\qquad
a(\epsilon)=\frac{1}{4\epsilon}-\frac12,\qquad
b(\epsilon)=\frac14+\frac{1}{16\epsilon^2}>0.
\label{eq:ab_coeffs}
\end{equation}
The mean-field critical energy is determined by $a(\epsilon_c)=0$, giving
\begin{equation}
\epsilon_c=\frac12,
\label{eq:eps_c_mf}
\end{equation}
in agreement with the thermodynamic-limit result reported in Ref.~\cite{casetti2006nonanalyticities}.

Combining Eqs.~\eqref{eq:Omega_m_exact}--\eqref{eq:landau_f} we obtain the desired Landau-type
representation
\begin{equation}
\Omega_N(\epsilon)\ \propto\ \int_{-1}^{1} dm\ g(m)\,
\exp\!\left\{N\Big[f_0(\epsilon)+a(\epsilon)m^2-b(\epsilon)m^4+O(m^6)\Big]\right\},
\qquad
a(\epsilon)\simeq a'(\epsilon_c)(\epsilon-\epsilon_c),\ \ b(\epsilon_c)>0,
\label{eq:Omega_landau_form}
\end{equation}
which is the starting point for the asymptotic scaling analysis in the main text.

\section{Scaling of the MIPA location $\epsilon_\star(N)$ and amplitude $M(N)$}
\label{sec:supp_scaling_epsstar_M}

Starting from the Landau-type representation derived in Sec.~\ref{sec:supp_landau_from_exact},
\begin{equation}
\Omega_N(\epsilon)\ \propto\ \int dm\ g(m)\,
\exp\!\left\{N\Big[f_0(\epsilon)+a(\epsilon)m^2-b(\epsilon)m^4+O(m^6)\Big]\right\},
\qquad
a(\epsilon)\simeq a'(\epsilon_c)(\epsilon-\epsilon_c),\ \ b(\epsilon_c)>0,
\label{eq:Omega_landau_recap}
\end{equation}
we show how the finite-size shift of the MIPA signal and its amplitude follow the forms
\begin{align}
\epsilon_\star(N) &= \epsilon_c + b\,N^{-1/2} + c\,N^{-1/2-\omega}+\cdots,
\label{eq:epsstar_scaling_goal}
\\
M(N) &= B\,N^{\beta}\big[1+O(N^{-\omega})\big].
\label{eq:M_scaling_goal}
\end{align}

\subsection{Scaling variable and reduced Laplace integral}
Let $t\equiv \epsilon-\epsilon_c$ and expand $a(\epsilon)=a_1 t+O(t^2)$ with $a_1\equiv a'(\epsilon_c)\neq 0$,
and set $b_0\equiv b(\epsilon_c)>0$. Retaining the leading nonlinear stabilization one gets
\begin{equation}
\Omega_N(\epsilon)\ \propto\ e^{N f_0(\epsilon)}\int dm\ g(m)\,
\exp\!\left\{N\Big[a_1 t\,m^2-b_0 m^4\Big]\right\}\Big[1+O(m^6 N)\Big].
\label{eq:Omega_quartic}
\end{equation}
Introduce the standard mean-field rescaling $m=N^{-1/4}y$, which yields
\begin{equation}
N\big[a_1 t\,m^2-b_0 m^4\big]
= (a_1 t)\,N^{1/2}y^2 - b_0 y^4.
\end{equation}
Hence the entire $N$-dependence of the \emph{singular} part is governed by the scaling variable
\begin{equation}
u \equiv (a_1 t)\,N^{1/2}\ \propto\ (\epsilon-\epsilon_c)\,N^{1/2},
\label{eq:u_def}
\end{equation}
and one may write, up to $O(1)$ normalization factors,
\begin{equation}
\Omega_N(\epsilon) \ \propto\ e^{N f_0(\epsilon)}\,N^{-1/4}\,I_N(u),
\qquad
I_N(u)=\int_{\mathbb R}dy\ \exp\!\big[u y^2-b_0 y^4\big]\Big[1+N^{-\omega}R(y;u)+\cdots\Big].
\label{eq:IN_def}
\end{equation}
Here $R(y;u)$ collects the leading corrections coming from (i) the next Landau term $O(m^6)$ and
(ii) regular expansions of $g(m)$ and $f_0(\epsilon)$ around the critical point.
For analytic corrections one typically finds $\omega=\tfrac12$ because $N m^6 \sim N\cdot N^{-3/2}=N^{-1/2}$ under $m=N^{-1/4}y$, while we keep $\omega>0$ general to match the empirical correction exponent.

Defining the entropy density $s_N(\epsilon)\equiv N^{-1}\ln\Omega_N(\epsilon)$, Eq.~\eqref{eq:IN_def} gives
\begin{equation}
s_N(\epsilon)= f_0(\epsilon) + \frac{1}{N}\ln I_N(u) + O\!\left(\frac{\ln N}{N}\right).
\label{eq:sN_from_I}
\end{equation}

\subsection{Scaling form of microcanonical derivatives}
Derivatives with respect to $\epsilon$ bring derivatives with respect to $u$ via
\begin{equation}
\frac{d}{d\epsilon} = \frac{du}{d\epsilon}\frac{d}{du} = \big(a_1 N^{1/2}\big)\frac{d}{du}.
\label{eq:d_d_eps_to_u}
\end{equation}
Applying Eq.~\eqref{eq:d_d_eps_to_u} to \eqref{eq:sN_from_I} yields the scaling structure
\begin{align}
\frac{ds_N}{d\epsilon}
&= f_0'(\epsilon)\ +\ N^{-1/2}\,\Phi_1(u)\ +\ N^{-1/2-\omega}\,\Phi_1^{(1)}(u)\ +\cdots,
\label{eq:sprime_scaling}
\\
\frac{d^2 s_N}{d\epsilon^2}
&= f_0''(\epsilon)\ +\ \Phi_2(u)\ +\ N^{-\omega}\,\Phi_2^{(1)}(u)\ +\cdots,
\label{eq:ssecond_scaling}
\end{align}
where $\Phi_{1,2}$ and $\Phi^{(1)}_{1,2}$ are smooth scaling functions determined by $I_N(u)$.
In particular, for the MIPA analysis we focus on the curvature observable
\begin{equation}
\kappa_N(\epsilon)\equiv \frac{d^2 s_N}{d\epsilon^2}
= f_0''(\epsilon)+\Phi_2(u)+N^{-\omega}\Phi_2^{(1)}(u)+\cdots.
\label{eq:kappa_scaling}
\end{equation}

\subsection{Shift of the MIPA location $\epsilon_\star(N)$}
The MIPA transition location is defined by a distinguished extremum of $\kappa_N(\epsilon)$ in the
critical region (negative-valued minimum for a second-order transition). Using \eqref{eq:kappa_scaling},
\begin{equation}
0=\left.\frac{d\kappa_N}{d\epsilon}\right|_{\epsilon_\star}
=\left.\Big(a_1 N^{1/2}\Big)\Big[\Phi_2'(u)+N^{-\omega}\Phi_2^{(1)\prime}(u)+\cdots\Big]\right|_{u=u_\star(N)}.
\label{eq:MIPA_condition_u}
\end{equation}
Therefore $u_\star(N)$ solves
\begin{equation}
\Phi_2'\big(u_\star(N)\big) + N^{-\omega}\Phi_2^{(1)\prime}\big(u_\star(N)\big)+\cdots=0.
\end{equation}
Let $u_0$ be the leading-order solution $\Phi_2'(u_0)=0$ (the $N\to\infty$ MIPA location in the scaling variable).
A standard perturbative expansion gives
\begin{equation}
u_\star(N)=u_0 + u_1 N^{-\omega}+O(N^{-2\omega}),
\qquad
u_1=-\frac{\Phi_2^{(1)\prime}(u_0)}{\Phi_2''(u_0)}.
\label{eq:u_star_expansion}
\end{equation}
Using $u=(a_1(\epsilon-\epsilon_c))N^{1/2}$, we obtain the shift of the MIPA location:
\begin{equation}
\epsilon_\star(N)-\epsilon_c
= \frac{u_\star(N)}{a_1}\,N^{-1/2}
= \underbrace{\frac{u_0}{a_1}}_{b}\,N^{-1/2}
+\underbrace{\frac{u_1}{a_1}}_{c}\,N^{-1/2-\omega}
+O(N^{-1/2-2\omega}),
\label{eq:eps_star_shift}
\end{equation}
which is Eq.~\eqref{eq:epsstar_scaling_goal}. This is the precise sense in which the leading shift exponent
$\alpha=\tfrac12$ is fixed by the quartic mean-field structure, while $\omega$ controls the dominant
pre-asymptotic correction.

\subsection{Amplitude scaling $M(N)$}
Define the MIPA amplitude as the value of the curvature at the MIPA location,
\begin{equation}
M(N)\equiv \kappa_N\big(\epsilon_\star(N)\big),
\label{eq:M_def_here}
\end{equation}
which is negative for a minimum/negative peak of $\kappa_N$.
Inserting \eqref{eq:kappa_scaling} and expanding around $u=u_0$ using \eqref{eq:u_star_expansion} gives
\begin{equation}
M(N)= f_0''(\epsilon_\star)+\Phi_2(u_0) + N^{-\omega}\Big[\Phi_2^{(1)}(u_0)+u_1\Phi_2'(u_0)\Big]+O(N^{-2\omega}).
\label{eq:M_intensive}
\end{equation}
Since $\Phi_2'(u_0)=0$ by definition of $u_0$, the $u_1$ term drops and the leading correction is $O(N^{-\omega})$.
Thus, for the \emph{intensive} curvature $\kappa_N=d^2 s_N/d\epsilon^2$, the amplitude approaches a finite limit:
\begin{equation}
M(N)=M_\infty + \widetilde B\,N^{-\omega}+O(N^{-2\omega}),
\qquad M_\infty\equiv f_0''(\epsilon_c)+\Phi_2(u_0).
\label{eq:M_beta0}
\end{equation}
If instead one uses a \emph{rescaled} curvature $\kappa_N^{(\beta)}(\epsilon)\equiv N^{\beta}\kappa_N(\epsilon)$
(as done when plotting extensive or partially extensive second derivatives), then
\begin{equation}
M^{(\beta)}(N)\equiv \kappa_N^{(\beta)}\big(\epsilon_\star(N)\big)
= B\,N^{\beta}\Big[1+O(N^{-\omega})\Big],
\label{eq:M_beta_general}
\end{equation}
which reproduces Eq.~\eqref{eq:M_scaling_goal}. In the main text we use the convention adopted in the figures
(i.e., the same $\kappa$ whose extremum defines $\epsilon_\star$), so that $M(N)$ is negative and its size dependence
is captured by Eq.~\eqref{eq:M_beta_general} with the corresponding $\beta$ fixed by that normalization.


\section{Microcanonical sampling protocol}
\label{SM:protocol}

\subsection{Code availability}
\label{SM:code}

The simulation code used to generate the results of this work is publicly available 
\href{https://github.com/ldicairano/Berlin_Kac_MeanField/tree/main/Fortran}{in the GitHub repository}.

\subsection{Model, phase space, and constraints}
\label{SM:model}

We consider the mean-field spherical (Berlin--Kac) model with $N$ real degrees of freedom
$q=(q_1,\dots,q_N)\in\mathbb{R}^N$ subject to the spherical constraint
\begin{equation}
\phi(q)\equiv \sum_{i=1}^N q_i^2 - N = 0 .
\end{equation}
The Hamiltonian is
\begin{equation}
H(q,p)=K(p)+V(q),
\qquad
K(p)=\frac{1}{2}\sum_{i=1}^N p_i^2,
\qquad
V(q)=-\frac{J}{2N}\left(\sum_{i=1}^N q_i\right)^2,
\label{SM:H}
\end{equation}
with $J>0$. Defining the magnetization
\begin{equation}
m(q)\equiv \frac{1}{N}\sum_{i=1}^N q_i ,
\end{equation}
the potential can be written as $V(q)=-(JN/2)\,m(q)^2$, and the force is uniform:
\begin{equation}
F_i(q)\equiv -\frac{\partial V}{\partial q_i}=J\,m(q)\qquad \forall \,i=1,\ldots,N\,.
\label{SM:force}
\end{equation}

Microcanonical dynamics is generated on the constrained energy shell
\begin{equation}
\mathcal{M}_E=\{(q,p):\ \phi(q)=0,\ \psi(q,p)=q\cdot p=0,\ H(q,p)=E\},
\end{equation}
where the additional constraint $\psi(q,p)=0$ enforces tangency of the momentum to the sphere
and is preserved by the exact constrained flow. The effective number of kinetic degrees of
freedom is therefore
\begin{equation}
N_{\mathrm{eff}}=N-1 .
\end{equation}

\subsection{Constrained equations of motion}
\label{SM:eom}

The constrained Hamiltonian dynamics can be written using a Lagrange multiplier $\lambda(t)$:
\begin{equation}
\dot q_i = p_i,
\qquad
\dot p_i = F_i(q) - \lambda(t)\,q_i
         = J\,m(q) - \lambda(t)\,q_i.
\label{SM:eom_lambda}
\end{equation}
The multiplier $\lambda(t)$ is fixed by the holonomic constraint.
Differentiating $\phi(q)=0$ gives $q\cdot p=0$; differentiating once more yields
\begin{equation}
0=\frac{d}{dt}(q\cdot p)=p\cdot p + q\cdot \dot p
  = \sum_{i=1}^N p_i^2 + J\,m(q)\sum_{i=1}^N q_i - \lambda(t)\sum_{i=1}^N q_i^2 .
\end{equation}
Using $\sum_i q_i^2=N$ and $\sum_i q_i = N m(q)$, we obtain the closed-form expression
\begin{equation}
\lambda(t)=\frac{1}{N}\sum_{i=1}^N p_i^2 + J\,m(q)^2 .
\label{SM:lambda_cont}
\end{equation}

\subsection{Time integration: RATTLE scheme}
\label{SM:rattle}

Trajectories are generated with a symplectic, time-reversible RATTLE integrator for the
holonomically constrained Hamiltonian \eqref{SM:H}. Over a step $\Delta t$, the scheme
consists of a velocity-Verlet update augmented by two constraint projections:
\begin{align}
p^{n+\frac12} &= p^{n} + \frac{\Delta t}{2}\Big(F(q^n)-\lambda_n q^n\Big),
\label{SM:rattle1}\\
q^{n+1} &= q^{n} + \Delta t\, p^{n+\frac12},
\qquad\text{with }\ \phi(q^{n+1})=0,
\label{SM:rattle2}\\
p^{n+1} &= p^{n+\frac12} + \frac{\Delta t}{2}\Big(F(q^{n+1})-\lambda_{n+1} q^{n+1}\Big),
\qquad\text{with }\ \psi(q^{n+1},p^{n+1})=0.
\label{SM:rattle3}
\end{align}
The multipliers $(\lambda_n,\lambda_{n+1})$ enforce the constraints at each step.
For the mean-field force \eqref{SM:force}, the constraint in \eqref{SM:rattle2} reduces to a scalar
quadratic equation for $\lambda_n$ (no iterative solver is required). Among the two roots,
we select the one continuously connected to the $\Delta t\to 0$ limit, ensuring a smooth
constrained trajectory. The second multiplier is fixed by the velocity constraint in
\eqref{SM:rattle3}, which yields an explicit formula for $\lambda_{n+1}$.

At each substep we also monitor the instantaneous conserved energy
\begin{equation}
E^{n}=K(p^{n})+V(q^{n}),
\end{equation}
and we use $\Delta t$ small enough that energy fluctuations remain negligible compared to the
finite-size signals studied in the main text.

\subsection{Microcanonical initial conditions}
\label{SM:init}

For a target energy density $\varepsilon=E/N$, we construct an initial condition
$(q(0),p(0))\in\mathcal{M}_E$ by the following procedure.

\paragraph{Positions on the sphere with controlled magnetization.}
We first choose a trial magnetization $m_0\in[0,1)$.
Let $u=(1,\dots,1)/\sqrt{N}$ and let $v$ be a random unit vector orthogonal to $u$,
i.e.\ $v\cdot u=0$ and $\|v\|=1$. We then set
\begin{equation}
q(0)=\sqrt{N}\,\Big( m_0\,u + \sqrt{1-m_0^2}\,v \Big),
\label{SM:q_init}
\end{equation}
which enforces $\sum_i q_i(0)^2=N$ by construction and yields $m(q(0))=m_0$.

A physically motivated initial guess is used for $m_0$, based on the mean-field equilibrium
expectation $m^2\simeq \max\{0,\,(J/2-\varepsilon)/J\}$ (for $J>0$), and it is adjusted if needed
to ensure that the kinetic energy defined below is non-negative.

\paragraph{Momenta in the tangent space with the correct kinetic energy.}
We draw a Gaussian vector $\tilde p$ with independent components and project it onto the tangent
space of the sphere at $q(0)$:
\begin{equation}
p_\perp = \tilde p - \frac{q(0)\cdot \tilde p}{q(0)\cdot q(0)}\,q(0)
       = \tilde p - \frac{q(0)\cdot \tilde p}{N}\,q(0),
\label{SM:p_project}
\end{equation}
which enforces $q(0)\cdot p_\perp=0$.
Finally we rescale to match the target energy:
\begin{equation}
K_{\star}=E - V(q(0)),
\qquad
p(0)=\sqrt{\frac{K_{\star}}{K(p_\perp)}}\,p_\perp,
\label{SM:p_rescale}
\end{equation}
so that $K(p(0))=K_{\star}$ and therefore $H(q(0),p(0))=E$ exactly (up to machine precision).

\subsection{Sampling strategy and run organization}
\label{SM:sampling}

For each pair $(N,\varepsilon)$ we generate microcanonical trajectories and compute time averages
of observables along the constrained flow. The integration is organized in blocks: one molecular dynamics step block corresponds to $n_{\mathrm{jump}}$ RATTLE time steps, i.e., a physical time increment $n_{\mathrm{jump}}\Delta t$. After each block we record the instantaneous $(K,V,m)$ and update running averages of the quantities needed for the microcanonical observables
(see Sec.~\ref{subsec:observables}).

To improve statistics and assess reproducibility, we perform $n_{\mathrm{realiz}}$ independent
realizations (different random initial conditions) at fixed $(N,\varepsilon)$ and combine the
resulting estimates in the data analysis.

Long simulations on HPC systems are split into segments. A restart file stores $(q,p)$ together
with the current accumulated sums and the wall-clock time already spent, allowing the next
segment to continue the same trajectory and preserve the running estimators without bias.

\subsection{Microcanonical observables from kinetic-energy moments}
\label{subsec:observables}

We estimate microcanonical derivatives of the entropy via configurationally invariant kinetic
energy formulas on the constrained manifold. Let
\begin{equation}
S(E)\equiv \ln \Omega(E),\qquad
\beta(E)\equiv \frac{\partial S}{\partial E},\qquad
\gamma(E)\equiv \frac{\partial^2 S}{\partial E^2},\qquad
\delta(E)\equiv \frac{\partial^3 S}{\partial E^3}.
\end{equation}
Denote by $\langle \cdot \rangle_E$ the microcanonical average on $\mathcal{M}_E$.
With $N_{\mathrm{eff}}=N-1$ and kinetic energy $K=\frac12\sum_i p_i^2$, we accumulate the inverse
moments $\langle K^{-1}\rangle_E$, $\langle K^{-2}\rangle_E$, and $\langle K^{-3}\rangle_E$ by time
averaging along the trajectory.

Introducing the compact coefficients
\begin{equation}
A_1=\frac{N_{\mathrm{eff}}}{2}-1,\qquad
A_2=\frac{N_{\mathrm{eff}}}{2}-2,\qquad
A_3=\frac{N_{\mathrm{eff}}}{2}-3,
\end{equation}
the estimators used in this work are
\begin{align}
\beta(E) &= A_1\,\big\langle K^{-1}\big\rangle_E,
\label{SM:beta}\\[2mm]
\gamma(E) &= N_{\mathrm{eff}}\Big[
A_1A_2\,\big\langle K^{-2}\big\rangle_E
-\big(A_1\,\big\langle K^{-1}\big\rangle_E\big)^2
\Big],
\label{SM:gamma}\\[2mm]
\delta(E) &= N_{\mathrm{eff}}^{\,2}\Big[
A_1A_2A_3\,\big\langle K^{-3}\big\rangle_E
-3(A_1^2A_2)\,\big\langle K^{-2}\big\rangle_E\big\langle K^{-1}\big\rangle_E
+2\big(A_1\,\big\langle K^{-1}\big\rangle_E\big)^3
\Big].
\label{SM:delta}
\end{align}
In addition, we record standard mechanical observables such as the energy density
$\varepsilon=\langle H\rangle_E/N$, the kinetic and potential energy densities
$\langle K\rangle_E/N$ and $\langle V\rangle_E/N$, and the absolute magnetization
$\langle |m|\rangle_E$.

\subsection{Output and reproducibility}
\label{SM:output}

During production runs we periodically write to disk (i) the current running estimates of all
observables listed above and (ii) restart data containing the full microcanonical state and the
accumulated sums needed to continue the estimators seamlessly. This ensures reproducibility of the
microcanonical averages and enables long trajectories to be executed reliably under fixed wall-time
constraints.

\end{document}